# Multi-head Watson- Crick quantum finite automata

Debayan Ganguly[1], Kingshuk Chatterjee[2], Kumar Sankar Ray[3]

[1] Government College of Engineering and Leather Technology, Block-LB, Sector –III, Kolkata-700106

[2] Government College of Engineering and Ceramic technology, Kolkata-700010

[3] Electronics and Communication Sciences Unit, Indian Statistical Institute, Kolkata-700108

debayan3737@gmail.com[1], kingshukchaterjee@gmail.com[2], ksray@isical.ac.in[3]

**Abstract.** Watson-Crick quantum finite automata were introduced by Ganguly et.al. by combining properties of DNA and Quantum automata. In this paper we introduce a multi-head version of the above automaton. We further show that the multi-head variant is computationally more powerful than one-way multi-head reversible finite automata. In fact we also show that the multi-head variant accepts a language which is not accepted by any one-way multi-head deterministic finite automata

## 1 Introduction

 Once Bennet [1] showed that reversible Turing machine has the same computational power as classical Turing machine, the interest in reversibility of automata peaked. Morita [2] worked on two-way variant of the multi-head reversible finite automata and showed that they are equivalent in computational power to multi-head deterministic finite automata. Kutrib et.al.[3] explored the one-way variant of the above model. The interest in reversible automata comes from its lossless nature of information processing. Quantum finite automata are inherently reversible due to reversible nature of quantum operations. But the reversible nature significantly reduces the computing power of one way quantum finite automata [4] and it accepts a subset of regular languages. In order to increase the computing power of quantum automata Ganguly et.al. [5] introduced properties of Watson-Crick automata [6] in quantum automata and showed that it significantly increased the power of quantum automata. In this paper we introduce a multi-head variant of the Watson-Crick quantum automata defined by Ganguly et. al.

The main claims of this paper are as follows:

- multi-head variant of Watson-Crick quantum automata is computationally more powerful than one-way multi-head reversible finite automata.
- multi-head variant of Watson-Crick quantum automata accepts a language which is not accepted by any one-way multi-head deterministic finite automata

## 2  Preliminaries and Definitions

The symbol V denotes a finite alphabet. The set of all finite words over V is denoted by $V^*$, which includes the empty word $\lambda$. The set of all non-empty words over the alphabet V is denoted by $V^+=V^*-\{\lambda\}$. For $w \in V^*$, the length of w is denoted by $|w|$. Let $u \in V^*$ and $v \in V^*$ be two words and if there is some word $x \in V^*$, such that $v=ux$, then u is a prefix of v, denoted by $u \leq v$. Two words, u and v are prefix comparable denoted by $u \sim_p v$ if u is a prefix of v or vice versa.

**Watson-Crick automata**

A Watson-Crick automaton is a 6-tuple of the form $M=(V,\rho,Q,q_0,F,\delta)$ where V is a set of alphabet. The set of states is denoted by Q, $\rho \subseteq V \times V$ denotes a complementarity relation. $q_0$ is the initial state and $F \subseteq Q$ is the set of final states. The function $\delta$ contains a finite number of transition rules of the form $q\binom{w_1}{w_2} \to q'$, which denotes that the automaton in state q parses $w_1$ in upper strand and $w_2$ in lower strand and goes to state q' where $w_1, w_2 \in V^*$. The symbol $\begin{bmatrix}w_1\\w_2\end{bmatrix}$ is different from $\binom{w_1}{w_2}$. While $\binom{w_1}{w_2}$ is just a pair of strings written in that form instead of $(w_1,w_2)$, the symbol $\begin{bmatrix}w_1\\w_2\end{bmatrix}$ denotes that the two strands are of same length i.e. $|w_1|=|w_2|$ and the corresponding symbols in two strands are complementarity in the sense given by the relation $\rho$. The notation $\begin{bmatrix}V\\V\end{bmatrix}_\rho = \{\begin{bmatrix}a\\b\end{bmatrix} \mid a,b \in V, (a,b) \in \rho\}$ and $WK_\rho(V) = \begin{bmatrix}V\\V\end{bmatrix}_\rho^*$ denotes the Watson-Crick domain associated with V and $\rho$.

A transition in a Watson-Crick finite automaton can be defined as follows:

For $\binom{x_1}{x_2}, \binom{u_1}{u_2}, \binom{w_1}{w_2} \in \binom{V^*}{V^*}$ such that $\begin{bmatrix}x_1 u_1 w_1\\x_2 u_2 w_2\end{bmatrix} \in WK_\rho(V)$ and $q,q' \in Q$, $\binom{x_1}{x_2}q\binom{u_1}{u_2}\binom{w_1}{w_2} \Rightarrow \binom{x_1}{x_2}\binom{u_1}{u_2}q'\binom{w_1}{w_2}$ iff there is transition rule $q\binom{u_1}{u_2} \to q'$ in $\delta$ and $\Rightarrow^*$ denotes the transitive and reflexive closure of $\Rightarrow$. The language accepted by a Watson-Crick automaton M is $L(M)=\{w_1 \in V^* | q_0\begin{bmatrix}w_1\\w_2\end{bmatrix} \Rightarrow^* \begin{bmatrix}\lambda\\\lambda\end{bmatrix}q$, with $q \in F$, $w_2 \in V^*, \begin{bmatrix}w_1\\w_2\end{bmatrix} \in WK_\rho(V)\}$.

**Multi-head reversible finite automata**

The following definition of one-way reversible multi-head finite automata and deterministic multi-head finite automata are obtained from Kutrib et.al. [**Error! Reference source not found.**] . In order to define one-way reversible multi-head finite automata

Kutrib also defined one-way deterministic multi-head finite automata. We will also do so in order to define the reversible multi-head automata in the same manner as Kutrib.

A one-way deterministic k head finite automaton (1DFA(k)) is a system M= (Q, V, k, #, $, $q_0$, F, δ) where

1) Q is the finite set of states.
2) V is the finite set of input symbols.
3) k≥1 is the number of heads.
4) #∉V is the left end marker.
5) $∉V is the right end marker.
6) $q_0$∈Q is the initial state.
7) F⊆Q is the set of accepting states and
8) δ:Q×(V∪{#,$})$^k$→Q×{0,1}$^k$ is the partial transition function where 1 means to move the head one position to the right and 0 means to keep the head in the current position. Whenever δ(q',($a_1,a_2,…,a_k$))=(q,($d_1,d_2,…,d_k$)) is defined then $d_i$=0 if $a_i$=$, 1≤i≤k.

The configuration of a 1DFA(k) M= (Q, V, k, #, $, $q_0$, F, δ) at some time t≥0 is a triple $c_t$=(w,q,P) where w∈V$^*$, q∈Q is the current state and P=($p_1$, $p_2$,…,$p_k$)∈{0,1,…,|w|+1}$^k$ gives the current head position. If $p_i$=0, then the i$^{th}$ head is scanning the symbol #. If it satisfies 1≤$p_i$≤|w|, then the i$^{th}$ head is scanning the $p_i$th symbol of w and if $p_i$=|w|+1 then the i$^{th}$ head is scanning the end marker symbol $ at the end of w. The initial configuration for input is set to (w, $q_0$, (0,...,0)) where string in the input tape is of the form  #w$. During its course of computation M goes through a sequence of configurations. One step from a configuration to its successor configuration is denoted by ⊢. Let w=$x_1x_2…x_n$, $x_0$=# and $x_{n+1}$=$, we write (w,q,($p_1,p_2,...,p_k$))⊢(w,q',($p_1+d_1,p_2+d_2,…,p_k+d_k$)) if and only if δ(q,($x_{p_1}, x_{p_2},…, x_{p_k}$))=(q',($d_1,d_2,…,d_k$)) exists. The reflexive and transitive closure of ⊢ is denoted by ⊢$^*$. The language accepted by 1DFA(k) is precisely the set of words w such that there is some computation beginning with #w$ and the 1DFA(k) halts in an accepting state. A 1DFA(k) halts when the transition function is not defined for the current situation.

L(M)={w∈V$^*$|(w,$q_0$,(0,…,0))⊢$^*$(w,q,($p_1,p_2,…,p_k$)), q∈F, and M halts in (w,q,($p_1,p_2,…,p_k$))}.

Based on definition of 1DFA(k) we define one-way multi-head reversible finite automaton in the following manner:

Let M be a 1DFA(k) and C be the set of all reachable configurations that occur in any computation of M beginning with an initial configuration and (w,q,($p_1,p_2,…,p_k$))∈C with w=$x_1x_2,…,x_n$, $x_0$=# and $x_{n+1}$=$.

M is said to be reversible, if the following two conditions are fulfilled:

1) For any two transitions δ(q',($a_1,a_2,...,a_k$))=(q,($d_1,d_2,...d_k$)) and δ(q'',($a_1',a_2',...,a_k'$))=(q,($d_1',d_2',...,d_k'$)) it holds that ($d_1,d_2,...d_k$)= ($d_1',d_2',...,d_k'$).

2) There is at most one transition of the form
δ(q',($x_{p_1-d_1}, x_{p_2-d_2},…, x_{p_k-d_k}$))=(q,($d_1,d_2,...,d_k$)).

**One-way quantum finite automata**

One-way quantum finite automaton is a eight tuple $M=(Q,V,\delta,q_0,Q_{acc},Q_{rej},\#,\$)$ where Q is a finite set of states, V is the input alphabet, $\delta$ is the transition function, $q_0 \in Q$ is the initial state and $Q_{acc} \subset Q$, and $Q_{rej} \subset Q$ are sets of accepting and rejecting states. The states in $Q_{acc}$ and $Q_{rej}$ are called halting states and the states in $Q_{non}=Q-(Q_{acc} \cup Q_{rej})$ are called the non-halting states. '#' and '$' are symbols that do not belong to V. We use '#' and '$' as the left and right end-markers respectively. The working alphabet of M is $\Gamma=V\cup\{\#,\$\}$. A superposition of an automaton M is an element in the Hilbert space $l_2(Q)$ (the space of mappings from Q to $\mathbb{C}$ with $l_2$ norm). We will use $\Psi$ to denote elements of $l_2(Q)$. For $q \in Q$, $|q\rangle$ denotes the unit vector with value 1 at q and 0 elsewhere i.e $|q\rangle =1.q$. All elements of $l_2(Q)$ can be expressed as a linear combination of vectors $|q\rangle$. We will use $\psi$ to denote $l_2(Q)$. The transition function $\delta$ maps $Q \times \Gamma \times Q$ to $\mathbb{C}$ where $\mathbb{C}$ denotes the set of complex numbers. The value $\delta(q_1,a,q_2)$ is the amplitude of $|q_2\rangle$ in the superposition of states to which M goes from $|q_1\rangle$ after reading 'a'. For $a \in \Gamma$, $U_a$ is a linear transformation on $l_2(Q)$ defined by $U_a(|q_1\rangle)= \sum_{q_2 \in Q} \delta(q_1,a,q_2)|q_2\rangle$. We require all $U_a$ to be unitary. The computation of a one-way quantum finite automaton starts in the superposition $|q_0\rangle$. Then transformations corresponding to left end-marker '#', the letters of the input word w and the right end-marker '$' are applied. The transformation corresponding to $a \in \Gamma$ consists of two steps.

Let S be a nonempty subset of a vector space P. The span of S, denoted by span(S), is the set containing of all linear combinations of vectors in S.

1) First, $U_a$ is applied. The new superposition $\psi'$ is $U_a(\psi)$ where $\psi$ is the superposition before this step.

2) $\psi$ is observed with respect to the observable $E_{acc} \oplus E_{rej} \oplus E_{non}$ where $E_{acc}$=span $|q\rangle:q \in Q_a$, $E_{rej}$=span $|q\rangle:q \in Q_{rej}$, $E_{non}$=span $|q\rangle:q \in Q_{non}$. Here $\oplus$ denotes an orthogonal sum on a Hilbert space. $E_{acc}=\sum_{q_i \in Q_{acc}} \alpha_i|q_i\rangle$, $E_{rej} = \sum_{q_i \in Q_{rej}} \beta_i|q_i\rangle$, $E_{non}=\sum_{q_i \in Q_{non}} \gamma_i|q_i\rangle$. $\alpha_i, \beta_i, \gamma_i$ are the amplitudes associated with $q_i$ in superposition $|\Psi\rangle$. This observation gives $x \in E_i$ with probability equal to the amplitude of the projection of $\psi'$. After that the superposition collapses to the projection.

If we get $\psi' \in E_{acc}$, the input is accepted. If $\psi' \in E_{rej}$, the input is rejected. If $\psi' \in E_{non}$, the next transformation is applied. We regard these two transformations as reading a letter 'a'. The above stated definition of 1QFA is from [7]. For further clarity of the above mentioned definition and notations of 1QFA we may consider the definition in Section II and example in Section III of [7].

**Watson-Crick Quantum Finite Automata**

A Watson-Crick quantum finite automaton is a nine tuple $M=(Q,V,\delta,q_0,Q_{acc},Q_{rej},\rho,\#,\$)$ where Q is a finite set of states, V is the input alphabet, $\delta$ is the transition function, $q_0 \in Q$ is the initial state, $Q_{acc} \subset Q$ and $Q_{rej} \subset Q$ are sets of accepting and rejecting states. The complementarity relation $\rho$ is similar to Watson-Crick complementarity relation. The states in $Q_{acc}$ and $Q_{rej}$ are called halting states and the states in $Q_{non}=Q-(Q_{acc} \cup Q_{rej})$ are called the non-halting states. The symbols '#' and '$' do not belong to V. We use '#' and '$' as the left and right end-markers respectively. The working alphabet of M is $\Gamma=V \cup \{\#,\$\}$. The input tape is a double stranded input tape with two heads each on one of the strands where the letters in the corresponding positions on the input tape are according to the complementarity relation ρ. The word on the upper strand is accepted or rejected by the automaton. A superposition of M is any element of $l_2(Q)$. For $q \in Q$, $|q\rangle$ denotes the unit vector with value 1 at q and 0 elsewhere. All elements of $l_2(Q)$ can be expressed as a linear combination of vectors $|q\rangle$. We will use ψ to denote $l_2(Q)$. The transition function δ maps $Q \times \Gamma^2 \times Q \times \{0,1\}^2$ to $\mathbb{C}$ where $\mathbb{C}$ denotes the set of complex numbers. The value $\delta(q_1,a,b,q_2,d_1,d_2)$ is the amplitude of $|q_2\rangle$ in the superposition of states to which M goes from $|q_1\rangle$ after reading 'a' in the upper strand and 'b' in the lower strand and moving the upper head according to $d_1$ and lower head according to $d_2$ where zero denotes head stays in its position and one denotes head has moved to the right. For $a, b \in \Gamma$, $U_{a,b}$ is a linear transformation on $l_2(Q)$ defined by $U_{a,b}(|q_1\rangle) = \sum_{q_2 \in Q} \delta(q_1, a, b, q_2, d_1, d_2)|q_2\rangle$. We require all $U_{a,b}$ to be unitary. The check for well-formedness can be done in a similar manner as in [7] in the following way:

Consider the Hilbert space $l_2(Q)$, where Q is the set of internal states of the automaton M. A linear operator $U_{\sigma,\tau}: l_2(Q) \to l_2(Q)$ for each σ,τ pair and a function $D: Q \to \{0,1\}^2$ exist. The transition function δ is defined as

$$\delta(q,\sigma,\tau,q',d_1,d_2) = \begin{cases} \langle q' | U_{\sigma,\tau} | q \rangle & \text{if } D(q') = (d_1, d_2) \\ 0 & \text{if } D(q') \neq (d_1, d_2) \end{cases}$$

where $\langle q' | U_{\sigma,\tau} | q \rangle$ denotes the coefficient of $|q'\rangle$ in $U_{\sigma,\tau}|q\rangle$. M is well-formed if and only if $\sum_{q'} \overline{\langle q' | U_{\sigma,\tau} | q_1 \rangle} \langle q' | U_{\sigma,\tau} | q_2 \rangle = \begin{cases} 1 & \text{if } q_1 = q_2 \\ 0 & \text{if } q_1 \neq q_2 \end{cases}$ for each σ,τ pair. The condition mentioned is similar to the condition for reversibility in [7].

The input word w is of the form $\begin{bmatrix} w_1 \\ w_2 \end{bmatrix} \in WK_\rho(V)$, where the automaton accepts or rejects $w_1$ with some probability. Both strands begin with # and ends with $. The string #$w_1$$ is placed in the upper strand and #$w_2$$ in the lower strand.

Note that many values of $U_{\sigma,\tau}|q\rangle$ define transitions which we do not encounter during a computation of w for a particular M. We define those values arbitrarily in such a way that $U_{\sigma,\tau}$ is unitary. In general we specify only those values that matter for all other values. The automaton M goes to some state q where $q \in Q$. The informed assignment of the other values along with the reversibility condition ensures that the resulting operator is unitary. So for a state q if no value is mentioned for a pair σ,τ where σ,τ∈Γ, Γ is finite. Therefore number of such σ,τ pairs are also finite. $U_{\sigma,\tau}|q\rangle = |q_{rejq}\rangle$ where defining unmentioned transitions in this way ensures well-formed transitions. Moreover for a given automaton M if such a transition is employed it always rejects. This enables us to define automaton without mentioning all the σ,τ pairs. We assume these $q_{rejq}$'s belong to the set $Q_{rej}$ and $D(q_{rejq}) = (0,0)$. As these transitions are included in automaton by default when mentioning the set $Q_{rej}$ and Q we do not explicitly mention these $q_{rejq}$'s.

A Watson-Crick quantum finite automaton (WKQFA) is called a **strongly Watson-Crick quantum finite automaton** (SWKQFA) if the complementarity relation is identity. The following definition is in[5] .

## 3   Multi-head Watson-Crick quantum finite automata

A multi-head Watson-Crick quantum finite automaton is an eleven tuple M=(Q,V,δ,q₀,Q_acc,Q_rej,ρ,#,$,k₁,k₂) where Q is a finite set of states, V is the input alphabet, δ is the transition function, q₀∈Q is the initial state, Q_acc⊂Q and Q_rej⊂Q are sets of accepting and rejecting states. The complementarity relation ρ is similar to Watson-Crick complementarity relation. The states in Q_acc and Q_rej are called halting states and the states in Q_non=Q-(Q_acc∪ Q_rej) are called the non-halting states. The symbols '#' and '$' do not belong to V. We use '#' and '$' as the left and right end-markers respectively. The working alphabet of M is Γ=V ∪ {#,$}. The input tape is a double stranded input tape where the letters in the corresponding positions on the input tape are according to the complementarity relation ρ. The symbol $k_1$ denotes the number of heads on the upper strand and $k_2$ denotes the number of heads on the lower strand. The word on the upper strand is accepted or rejected by the automaton. A superposition of M is any element of $l_2(Q)$. For q∈Q, |q⟩ denotes the unit vector with value 1 at q and 0 elsewhere. All elements of $l_2(Q)$ can be expressed as a linear combination of vectors |q⟩. We will use ψ to denote $l_2(Q)$. The transition function δ maps $Q \times \Gamma^{k_1+k_2} \times Q \times \{0,1\}^{k_1+k_2}$ to ℂ where ℂ denotes the set of complex numbers. The value $\delta(q_1, a_1, a_2 \ldots a_{k_1} | b_1, b_2 \ldots b_{k_2}, q_2, d_1, d_2 \ldots d_{k_1} | p_1, p_2 \ldots p_{k_2})$ is the amplitude of |q₂⟩ in the superposition of states to which M goes from |q₁⟩ after reading $a_1, a_2 \ldots a_{k_1}$ using the $k_1$ heads in the upper strand and $b_1, b_2 \ldots b_{k_2}$ using the $k_2$ heads in the lower strand and moving the upper $k_1$ heads according to $d_1, d_2 \ldots d_{k_1}$ and lower $k_2$ heads according to $p_1, p_2 \ldots p_{k_2}$ where zero denotes head stays in its position and one denotes head has moved to the right. For $a_1, a_2 \ldots a_{k_1}, b_1, b_2 \ldots b_{k_2} \in \Gamma, U_{a_1,a_2\ldots a_{k_1}|b_1,b_2\ldots b_{k_2}}$ is a linear transformation on $l_2(Q)$ defined by

$$U_{a_1,a_2\ldots a_{k_1}|b_1,b_2\ldots b_{k_2}}(|q_1\rangle) = \sum_{q_2 \in Q} \delta(q_1, a_1, a_2 \ldots a_{k_1} | b_1, b_2 \ldots b_{k_2}, q_2, d_1, d_2 \ldots d_{k_1} | p_1, p_2 \ldots p_{k_2}) |q_2\rangle$$

We require all $U_{a_1,a_2\ldots a_{k_1}|b_1,b_2\ldots b_{k_2}}$ to be unitary. The check for well-formedness can be done in a similar manner as in [7] in the following way:

Consider the Hilbert space $l_2(Q)$, where Q is the set of internal states of the automaton M. A linear operator $U_{a_1,a_2\ldots a_{k_1}|b_1,b_2\ldots b_{k_2}}:l_2(Q) \to l_2(Q)$ for each $a_1, a_2 \ldots a_{k_1}|b_1, b_2 \ldots b_{k_2}$ combination and a function $D:Q \to \{0,1\}^{k_1+k_2}$ exist. The transition function δ is defined as

$$\delta(q_1, a_1, a_2 \ldots a_{k_1} | b_1, b_2 \ldots b_{k_2}, q_2, d_1, d_2 \ldots d_{k_1} | p_1, p_2 \ldots p_{k_2})$$
$$= \begin{cases} \langle q' | U_{a_1,a_2\ldots a_{k_1}|b_1,b_2\ldots b_{k_2}} | q \rangle & \text{if } D(q') = (d_1, d_2 \ldots d_{k_1} | p_1, p_2 \ldots p_{k_2}) \\ 0 & \text{if } D(q') \neq (d_1, d_2 \ldots d_{k_1} | p_1, p_2 \ldots p_{k_2}) \end{cases}$$

where $\langle q'|U_{a_1,a_2...a_{k_1}|b_1,b_2...b_{k_2}}|q\rangle$ denotes the coefficient of $|q'\rangle$ in $U_{a_1,a_2...a_{k_1}|b_1,b_2...b_{k_2}}|q\rangle$. M is well-formed if and only if $\sum_{q'} \overline{\langle q'|U_{a_1,a_2...a_{k_1}|b_1,b_2...b_{k_2}}|q_1\rangle} \langle q'|U_{a_1,a_2...a_{k_1}|b_1,b_2...b_{k_2}}|q_2\rangle = \begin{cases} 1 \text{ if } q_1 = q_2 \\ 0 \text{ if } q_1 \neq q_2 \end{cases}$ for each $a_1, a_2 ... a_{k_1}|b_1, b_2 ... b_{k_2}$ combination. The condition mentioned is similar to the condition for reversibility in [7].

The input word w is of the form $\begin{bmatrix} w_1 \\ w_2 \end{bmatrix} \in WK_\rho(V)$, where the automaton accepts or rejects $w_1$ with some probability. Both strands begin with # and ends with $. The string $\#w_1\$$ is placed in the upper strand and $\#w_2\$$ in the lower strand.

Note that many values of $U_{a_1,a_2...a_{k_1}|b_1,b_2...b_{k_2}}|q\rangle$ define transitions which we do not encounter during a computation of w for a particular M. We define those values arbitrarily in such a way that $U_{a_1,a_2...a_{k_1}|b_1,b_2...b_{k_2}}$ combination is unitary. In general we specify only those values that matter for all other values. The automaton M goes to some state q where q∈Q. The informed assignment of the other values along with the reversibility condition ensures that the resulting operator is unitary. So for a state q if no value is mentioned for $a_1, a_2 ... a_{k_1}|b_1, b_2 ... b_{k_2}$ combination where $a_1, a_2 ... a_{k_1}, b_1, b_2 ... b_{k_2} \in \Gamma$, $\Gamma$ is finite. Therefore number of such $a_1, a_2 ... a_{k_1}|b_1, b_2 ... b_{k_2}$ combinations are also finite. $U_{a_1,a_2...a_{k_1}|b_1,b_2...b_{k_2}}|q\rangle = |q_{rejq}\rangle$ where defining unmentioned transitions in this way ensures well-formed transitions. Moreover for a given automaton M if such a transition is employed it always rejects. This enables us to define automaton without mentioning all the $a_1, a_2 ... a_{k_1}|b_1, b_2 ... b_{k_2}$ combinations. We assume these $q_{rejq}$s' belong to the set $Q_{rej}$ and $D(q_{rejq})=(0,0)$. As these transitions are included in automaton by default when mentioning the set $Q_{rej}$ and Q, we do not explicitly mention these $q_{rejq}$s'.

A Multi-head Watson-Crick quantum finite automaton (WKQFA) is called a **strongly Multi-head Watson-Crick quantum finite automaton** (SMWKQFA) if the complementarity relation is identity.

## 4    Computational complexity of multi-head Watson-Crick quantum finite automata

In this section, we discuss the computational power of multi-head Watson-Crick quantum finite automata.

**Lemma 1:** Watson-Crick quantum finite automaton M=(Q,V,δ,$q_0$,$Q_{acc}$,$Q_{rej}$,ρ,#,$) has the same computational power as multi-head Watson-Crick quantum finite automaton M=(Q,V,δ,$q_0$,$Q_{acc}$,$Q_{rej}$,ρ,#,$,1,1).
**Proof:** Watson-Crick quantum finite automaton M=(Q,V,δ,$q_0$,$Q_{acc}$,$Q_{rej}$,ρ,#,$) is a model which has two heads one traversing the upper strand and the other traversing

the lower head, similarly multi-head Watson-Crick quantum finite automaton M=$(Q,V,\delta,q_0,Q_{acc},Q_{rej},\rho,\#,\$,1,1)$ is also a model where one head traverses the upper strand and another head traverses the lower strand, the transition functions and accepting conditions for both the model are same, Hence there computational power is also same.

**Theorem 1:** There is a strongly multi-head Watson-Crick quantum finite automaton with 1+1 heads that accepts the context sensitive language L = $\{ww|w \in \{a,b\}^*\}$.
**Proof:** From[5] we know that strongly Watson-Crick quantum finite automaton accepts the context sensitive language L = $\{ww|w \in \{a,b\}^*\}$. Also, from Lemma 1 we know that Watson-Crick quantum finite automaton M=$(Q,V,\delta,q_0,Q_{acc},Q_{rej},\rho,\#,\$)$ has the same computational power as multi-head Watson-Crick quantum finite automaton M=$(Q,V,\delta,q_0,Q_{acc},Q_{rej},\rho,\#,\$,1,1)$. Therefore from the above two statements we can conclude that there is a strongly Multi-head Watson-Crick quantum finite automaton with 1+1 heads that accepts the context sensitive language L = $\{ww|w \in \{a,b\}^*\}$.

**Theorem 2:** For every deterministic finite automaton which accepts a language L, we can find a multi-head Watson-Crick quantum finite automaton with 1+1 heads which accepts the same language L.
**Proof:** From [5] we know that for every deterministic finite automaton which accepts a language L we can find a Watson-Crick quantum finite automaton which accepts the same language L. Also, from Lemma 1 we know that Watson-Crick quantum finite automaton M=$(Q,V,\delta,q_0,Q_{acc},Q_{rej},\rho,\#,\$)$ has the same computational power as multi-head Watson-Crick quantum finite automaton M=$(Q,V,\delta,q_0,Q_{acc},Q_{rej},\rho,\#,\$,1,1)$. Therefore from the above two statements we can conclude that for every deterministic finite automaton which accepts a language L, we can find a multi-head Watson-Crick quantum finite automaton with 1+1 heads which accepts the same language L.

**Corollary 1:** Multi-head Watson-Crick quantum finite automaton with 1+1 heads can accept all regular languages.

**Proof:** From Theorem 2, for every deterministic finite automaton which accepts a language L, we can find a multi-head Watson-Crick quantum finite automaton with 1+1 heads which accepts the same language L. For every regular language there is a deterministic finite automaton which accepts that language, thus for every regular languages there is a Multi-head Watson-Crick quantum finite automaton with 1+1 heads that accepts it.

**Lemma 2:** The language
L=$\{\%w_1*x_1\%w_2*x_2...\%w_n*x_n | n \geq 0, w_i \in \{a,b\}^*, x_i \in \{a,b\}^*, \exists i \exists j: w_i=w_j, x_i \neq x_j\}$ is accepted by a multi-head quantum finite automaton with 1+1 heads and non-injective complementarity relation.
**Proof:** From [5] we know that there exists a Watson-Crick quantum finite automaton with non-injective complementarity relation which accepts the language L=$\{\%w_1*x_1\%w_2*x_2...\%w_n*x_n | n \geq 0, w_i \in \{a,b\}^*, x_i \in \{a,b\}^*, \exists i \exists j: w_i=w_j, x_i \neq x_j\}$, and as computational power of Watson-Crick quantum finite automaton and multi-head quantum finite automaton with 1+1 heads are same, therefore multi-head Watson-Crick automaton also accepts the language L.

**Lemma 3:** The language
L=$\{\%w_1{*}x_1\%w_2{*}x_2...\%w_n{*}x_n | n \geq 0, w_i \in \{a,b\}^*, x_i \in \{a,b\}^*, \exists i \exists j: w_i = w_j, x_i \neq x_j\}$ is not accepted by any deterministic multi-head finite automaton.
 The proof of Lemma 1 is in Yao et al. [8].

**Theorem 3:** The set of languages accepted by one-way reversible multi-head finite automata is a proper subset of set of languages accepted multi-head Watson-Crick quantum finite automata.

**Proof:** From the definition of strongly multi-head Watson-Crick quantum finite automata it is evident that if no superposition states are involved in the strongly multi-head Watson-Crick quantum finite automaton and due to the reversibility condition of quantum automata, then the automaton behaves like a one-way multi-head reversible finite automaton. Thus for every multi-head one-way reversible finite automaton there is a strongly Watson-Crick quantum finite automaton which accepts the same language by just imitating the transition and acceptance conditions of the multi-head reversible automata. Moreover, it has already been stated by Kutrib et al. [3] that set of languages accepted by multi-head reversible finite automata is a proper subset of set of languages accepted by multi-head deterministic finite automata. Thus there is no multi-head reversible finite automaton which accepts the language L=$\{\%w_1{*}x_1\%w_2{*}x_2...\%w_n{*}x_n | n \geq 0, w_i \in \{a,b\}^*, x_i \in \{a,b\}^*, \exists i \exists j: w_i = w_j, x_i \neq x_j\}$. But from Lemma 3, we see that multi-head Watson-Crick quantum automaton can accept the language L, which proves the Theorem.

## 5   Conclusion

In this paper, we have introduced a multi-head variant of the Watson-Crick quantum finite automata model. We were able to show that the multi-head variant is computationally more powerful than one-way reversible multi-head finite automata model. In fact it accepts languages which are not accepted by any one-way deterministic multi-head finite automata. In future we need to see the impact of unary languages on these models and undecidability questions associated with it.

## References

1. C. H. Bennett, "Logical reversibility of computation," *IBM Journal of Research and Development*, vol. 17, pp. 525–532, nov 1973.